# Computation of statistical power and sample size for *in vivo* research models

[1]Hasan Al-Nashash*^, [2]Jiajin Wei^, [3]Ke Yang^, [4]Ayman Alzaatreh,
[5,6]Mohsen Adeli*, [2]Tiejun Tong*, and [7]Angelo ALL*

**^** *Contributed equally - Shared co-first authorship*

**\*** *Corresponding authors*

[1]Hasan Awad Al-Nashash, Ph.D.
Affiliation: Department of Electrical Engineering, College of Engineering, American University of Sharjah, UAE
Institutional title: Professor of Electrical and Biomedical Engineering
Tel: (+971) 6515 2935
E-mail: hnashash@aus.edu

[2]Jiajin Wei
Affiliation: Department of Mathematics, Hong Kong Baptist University, Hong Kong
Institutional title: Ph.D. Student
Tel: (+852) 5703 6751
E-mail: 20482051@life.hkbu.edu.hk

[3]Ke Yang, Ph.D.
Affiliation: School of Statistics and Data Science, Beijing University of Technology, Beijing, China
Institutional title: Assistant Professor
Tel: (+86) 13520179020
E-mail: yangke@bjut.edu.cn

[4]Ayman Alzaatreh, Ph.D.
Affiliation: Department of Mathematics and Statistics, College of Arts and Sciences, American University of Sharjah, UAE
Institutional title: Associate Professor
Tel: (+971) 6 515 2916
Email: aalzaatreh@aus.edu

[5, 6]Mohsen Adeli, PhD
Institut für Chemie und Biochemie, Freie Universität Berlin, Takustr. 3, 14195 Berlin, Germany
Department of Chemistry, Lorestan University, Khorramabad, Iran
Institutional title: Professor
E-mail: aadeli@zedat.fu-berlin.de

[2]Tiejun Tong, Ph.D.
Affiliation: Department of Mathematics, Hong Kong Baptist University, Hong Kong
Institutional title: Professor
Tel: (+852) 3411 7340
E-mail: tongt@hkbu.edu.hk

[7]Angelo Homayoun All, MD, MBA
Affiliation: Department of Chemistry, Hong Kong Baptist University, Hong Kong
Institutional title: Associate Professor
Tel: (+852) 3411 5956
E-mail: angelo@hkbu.edu.hk
Address: Department of Chemistry, Sir Run Run Shaw Building, Room RRS844, Ho Sin Hang Campus, Faculty of Science, Hong Kong Baptist University, Hong Kong SAR, China

**Abstract**

Sample size calculation is crucial in biomedical *in vivo* research investigations mainly for two reasons: to design the most resource-efficient studies and to safeguard ethical issues when alive animals are subjects of testing. In this context, power analysis has been widely applied to compute the sample size by predetermining the desired statistical power and the significance level. To verify whether the assumption of a null hypothesis is true, repeated measures analysis of variance (ANOVA) is used to test the differences between multiple experimental groups and control group(s). In this article, we focus on the a priori power analysis, for testing multiple parameters and calculating the power of experimental designs, which is suitable to compute the sample size of trial groups in repeated measures ANOVA. We first describe repeated measures ANOVA and the statistical power from a practical aspect of biomedical research. Furthermore, we apply the G*Power software to conduct the a priori power analysis using examples of repeated measures ANOVA with three groups and five time points. We aim not to use the typical technically adapted statistical language. This will enable experimentalists to confidently formulate power calculation and sample size calculation easier and more accurately.

# I. Introduction

In biomedical research that uses in vivo model, sample size calculation plays an important role in experimental design aimed at assessing the safety, the effectiveness, the mechanism, and/or characterization of a given research inquiry that would achieve the objectives through a well-defined hypothesis. An experiment may not yield a correct assumption of the outcome(s), if the sample size is not adequate. This means if the sample size is insufficient, the test's power will be low and that is, the likelihood of making the correct decision is low, Thus, practically would be either non statistically significant results (incomplete study) or consuming resources with no true conclusion (non-efficient use of resources). On the other hand, a study with an excessively large sample size will not only waste the resources, but also bring ethical, moral, and legal issues related to unjustified causing pain, suffering, unnecessary euthanasia, and abuse of animals [1].

The process of calculating the sample size and power is referred to as power analysis [2], which has been extensively studied for various statistical models or tests of hypotheses. They include Student's *t*-tests, chi-square tests, *F* tests for regression models, and analysis of variance (ANOVA) [1, 3, 4]. Among different types of power analysis, the a priori power analysis is widely applied in various fields. Scientists often use the a priori power analysis to compute the sample size before the study by predetermining the desired statistical power, the significance level, and the effect size [1]. As an opposite, the post-hoc power analysis allows one to calculate the statistical power after completing the study, by providing the sample size, the significance level, and the effect size. Obviously, the resulting power does not help to establish an appropriate sample size before the study [3]. The minimum number of animals per group needs to be extrapolated based on power calculation in order to obtain statistically significant results that support meaningful conclusions. In this manuscript, our goal is to represent the a priori power analysis and propose to compute and determine the sample size, applicable to *in vivo* biomedical experimental design.

The remaining sections of the paper are organized as follows: in section II, we discuss data analysis using repeated measure ANOVA followed by multiple samples repeated measures ANOVA. More details on the theory of repeated measure ANOVA is provided in the supplementary. In section III, we describe the sample size calculations and how to use the computer software used to achieve that. Following sample size calculation, we will describe in section IV the *in vivo* experimental design and data collection with focus on spinal cord injury. In section V, elaborate application examples illustrating how to calculate the sample size and perform data analysis will be provided. Finally, in section VI, we will finalize the paper with discussion and conclusions.

## II. Data Analysis

**Repeated Measures ANOVA**

The 'repeated measures' design is basically one of the most used strategies that allows researchers to assess not only the differences among groups or factors but also the changes in effectiveness of a variable longitudinally (a series of time points over a relatively long period of time) [5, 6]. For instance, to investigate a particular mechanism [7-11], the safety and efficacy of the stem cell replacement therapy [12], therapeutic effects of inducing hypothermia post-neurotrauma [13], characterization of neurotrauma [14], presence, absence, and beneficial effects of neuronal plasticity [15, 16] or neuromodulation would need alive animals randomized in many experimental groups that must also be compared with the positive and negative control groups over a relatively long period of time and with many examination time points using various assessment tools. Needless to say that the quantity of consumables, resources, manpower, and necessary time to collect such large amount of raw data are inevitable parameters that must be considered. In these examples, one could appreciate the true value of the sample size (minimal number of animals per groups that could provide acceptable power and meaningful results to draw conclusive presumption).

By applying repeated measures ANOVA, one can verify the quantitative and qualitative differences among multiple assessments in each of the experimental groups and compare them to the control group(s) over long-time. It is also noteworthy that the a priori power analysis should be first conducted to determine the number of subjects (or sample size) needed for the study with repeated measures. In the supplementary section, we describe the theory of repeated measures ANOVA that specifies the three common tests of hypotheses, provide the formulas necessary for calculating the sample size and power, and then use the G*Power software to conduct the a priori power analysis with an illustrative example [3, 17]. Repeated measures ANOVA has been extensively studied [18, 19].

In a study with repeated measures, we let $g$ be the number of groups and $n_j$ be the number of subjects in group *j*. Consequently, there are a total of $n = \sum_{j=1}^{g} n_j$ subjects, which is also referred to as the total sample size. For each of the subjects, we further take repeated measures at $t$ time points. For illustration, **Table 1** depicts the layout of repeated measures data.

**Table 1:** Layout for multiple-sample repeated measures data.

| Group | Subject | Measurements order | | | |
|---|---|---|---|---|---|
| | | Time 1 | Time 2 | … | Time t |
| 1 | 1 | $y_{111}$ | $y_{112}$ | … | $y_{11t}$ |
| | 2 | $y_{121}$ | $y_{122}$ | … | $y_{12t}$ |
| | ... | ... | | | |
| | $n_1$ | $y_{1 n_1 1}$ | $y_{1 n_1 2}$ | … | $y_{1 n_1 t}$ |
| … | 1 | … | … | … | … |
| | 2 | | | | |
| | ... | … | … | … | … |
| | $n_2$ | … | … | … | … |
| g | 1 | $y_{g11}$ | $y_{g12}$ | … | $y_{g1t}$ |
| | 2 | $y_{g21}$ | $y_{g22}$ | … | $y_{g2t}$ |
| | ... | … | | | |
| | $n_g$ | $y_{g n_g 1}$ | $y_{g n_g 2}$ | … | $y_{g n_g t}$ |

When $g=1$, **Table 1** depicts the layout of one-sample repeated measures and when $g>1$, it depicts the layout of multiple-sample repeated measures data. The response variable $y_{kij}$ is assumed to have a normal distribution for each group $k=1,\ldots,g$.

The subjects are independent and are assigned to each group at random. Because of the nature of repeated measures, it is also assumed that the observations for each subject are correlated.

The variances between pairwise differences of repeated observations must be constant which is referred to as the sphericity assumption. With this sphericity assumption the model is similar to the randomized block design with random blocks. The subjects in this case are the blocks. For testing the differences among the $t$ time points, the null and alternative hypotheses are:

$H_0$ : the mean responses are all equal among time points,

$H_1$ : the mean responses are not all equal among time points.

**Multiple Samples Repeated Measures ANOVA**

Multiple samples occur when there are more than one group. For example, when comparing the control group to the experimental group with repeated measurements are taken over time.

For a multiple-sample problem, we can formulate them as three hypothesis testing problems as follows:

(i) For testing the differences among the $g$ groups, the null and alternative hypotheses are

$H_0$ : the mean responses are all equal among groups,

$H_1$ : the mean responses are not all equal among groups.

(ii) For testing the differences among the $t$ time points, the null and alternative hypotheses are

$H_0$ : the mean responses are all equal among time points,

$H_1$ : the mean responses are not all equal among time points.

(iii) For testing the interaction between group and time, the null and alternative hypotheses are

$H_0$ : there is no interaction between group and time,

$H_1$ : there is an interaction between group and time.

### III. Sample Size Calculation

To calculate the statistical power associated with the test in (S4), we let $F_{\lambda,g-1,n-g}$ denote the noncentral $F$ distribution with noncentral parameter $\lambda$ and degrees of freedom $g-1$ and $n-g$, then the statistical power of the test can be calculated as

$$\text{Power} = 1 - \beta = 1 - F_{\lambda,g-1,n-g}(F^{-1}_{g-1,n-g}(1-\alpha)), \qquad (1)$$

where $\beta$ is the probability of a type II error (probability of failing to reject $H_0$ while $H_1$ is true). Moreover, the noncentral parameter is given as $\lambda = f^2 tn/(1 + (t-1)\rho)$, where $f$ is the effect size and $\rho$ is the correlation among repeated measures. For more details on these two parameters, see **Table 2**.

Moreover, to calculate the sample size, the formula in (1) can also be employed with the notations remaining all the same. Specifically, when the values of $f$, $\rho$, $g$, $t$, $\alpha$ and $1-\beta$ are assigned, the sample size $n$ will be the only unknown parameter in the formula and so can be readily solved by a classical numerical algorithm.

For the hypothesis testing problems in (S5) and (S6), we note that the calculation of the sample size follows a similar procedure as that in problem (S4). In addition, for the time points, if the repeated measures have different variances and different correlation coefficients (non-sphericity case), the distributions of the test statistics under the null and alternative hypotheses need to be corrected by applying a correction $\varepsilon$, which is also an input parameter in the G*Power software [17]. For more details on the correction $\varepsilon$, see also **Table 2**.

### IV.

**Table 2.** More explanations on the 3 key parameters in repeated measures ANOVA, which are also specified as the input parameters the G*Power software.

| | |
|---|---|
| $f$ | The effect size $f$ reflects the amount of variation due to different treatments. Specifically for the hypothesis testing problem (i), $f$ is defined as a ratio of the between and within-group standard deviations. The between-group standard deviation is the average of the squared distances between the mean response in each group and the overall mean response. The within-group standard deviation |
| | is assumed to be identical across different groups in repeated measures ANOVA. Hence, a larger between-group standard deviation will lead to a larger effect size $f$, which reflects more significant treatment effects. In the a priori power analysis, both the between and within-group standard deviations are typically unknown before the study. In view of this, we can apply Cohen's $f$ values with $f = 0.1$, 0.25 and 0.4 being the small, medium and large effect sizes, respectively [2]. For calculating $f$ in the other two hypothesis testing problems, one may refer to the references [3,17]. |
| $\rho$ | $\rho$ is the population correlation coefficient among the repeated measures. Before the study, there is a lack of data for estimating $\rho$. In the literature, researchers often apply $\rho = 0.5$ to represent a medium positive correlation coefficient [3,17]. |
| $\varepsilon$ | For the hypothesis testing problems in (ii) and (iii), when we assume that the repeated measures have an identical variance, and each pair of the repeated measures has an identical correlation coefficient, this assumption is known as the sphericity assumption [19]. If the sphericity assumption does not hold, the distributions of the test statistics under the null and alternative hypotheses need to be corrected by applying a non-sphericity correction parameter $\varepsilon$ [17]. On the other hand, if the sphericity assumption is met, it yields $\varepsilon = 1$. When $\varepsilon$ is unknown in the a priori power analysis, we assume $\varepsilon = 1$ by following the literature [3]. |

**Computer Software**

Many readily available computer programs, such as SPSS, R, and SAS software, can be used to analyze the data with repeated measures. For example, the users can utilize PROC GLM in SAS software. A good tutorial on how to use PROC GLM in repeated measures is available in [26]. R users can use anova_test() function in the rstatix library to run repeated measures ANOVA.

**Sample Size Calculation and Power Analysis Using the G*Power Software**

For repeated measures data, we recommend the free software G*Power for power analysis and sample size calculation. The G*Power can simulate the power for various sample sizes. For example, the sample size required for repeated measures ANOVA is determined by the effect size (*f*), number of groups (*g*), number of measurements (*t*), the level of correlations between the repeated measures ($\rho$) and the type I error ($\alpha$). The sample size can be chosen based on the desired level of power (for example, 80%) and effect size (*f*). According to [27, pp. 285-287], effect sizes can range from small at *f*=0.1, to medium at *f*=0.25, to large at *f*=0.4 (see **Table 2**). A larger sample size is required to detect small effect sizes (see **Figures 1b, 1d, and 1f**). Moreover, there are existing literature that explains in some detail how to choose the sample size for repeated measures ANOVA using the G*Power software and presents tutorial of how to set up the effect size, power, significant level, correlations and other parameters in the G*Power [28, 29].

Next, we present an example of how to determine the required sample size for tests in (S4), (S5) and (S6) respectively.

**The Power Analysis**

In this example, we apply the G*Power software to calculate the required sample size using the a priori power analysis. We consider a repeated measures study with 4 groups and 5 time points. To conduct the hypothesis testing in (S), we select "F tests" as the test family and also specify "ANOVA: Repeated measures, between factors" as the statistical test. Next, to calculate the sample size, we choose "A priori: Compute required sample size- given α, power, and effect size" as the type of power analysis. Finally, for the input parameters, we assign $g = 4$ for the number of groups, $t = 5$ for the number of repeated measures, $\alpha = 0.05$ for the significance level, and $1 - \beta = 0.8$ as the desired statistical power. Note that the remaining two parameters, including $f$ and $\rho$, are not so obvious, and in many cases, need the relevant expertise in the field of study. In case there is no prior knowledge, **Table 2** can be followed so that we assign $f = 0.25$ as the medium effect size and $\rho = 0.5$ as the medium correlation among repeated measures. Finally, we report the analytical results from the G*Power in **Figure**

**1a**, which shows that a minimum sample size of 112 is required to yield a statistical power of 0.8 or greater.

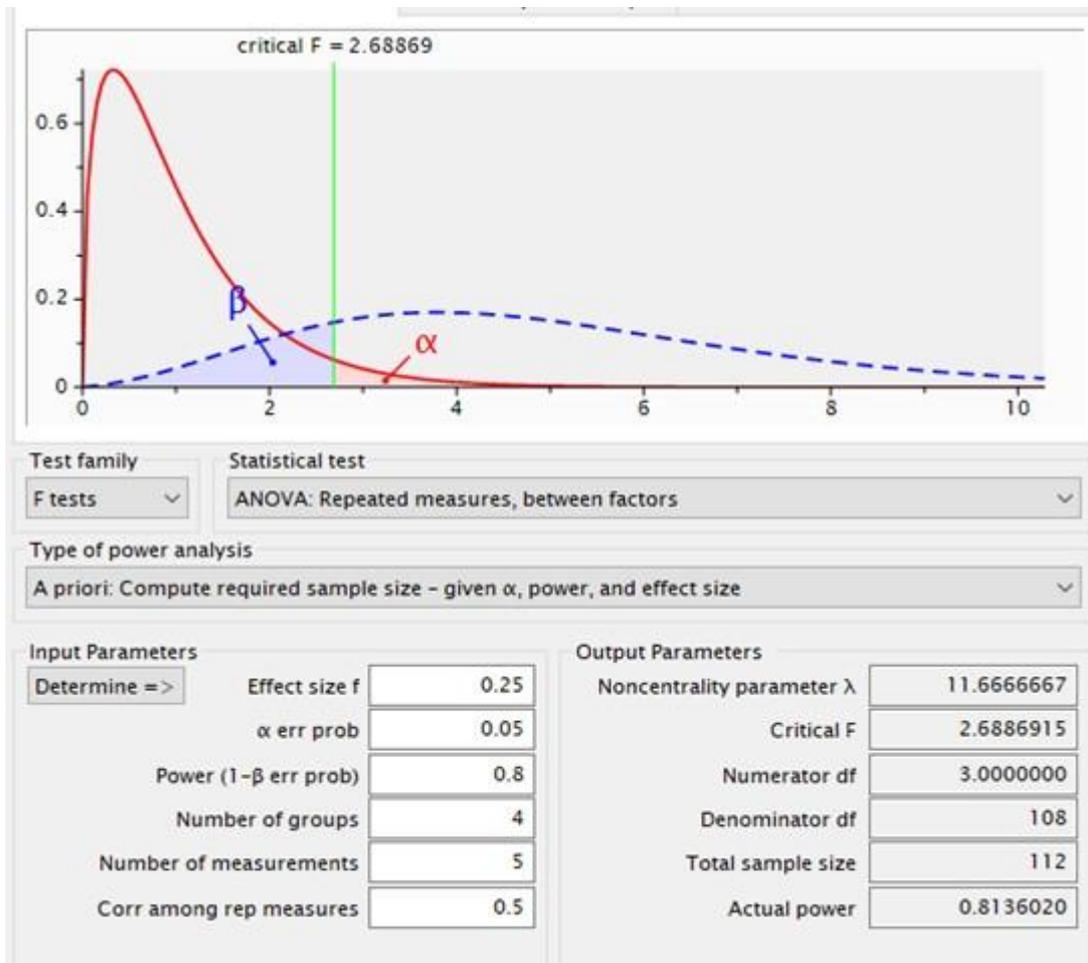

**Fig. 1a:** The a priori power analysis to calculate the sample size using the G*Power for testing the differences among groups.

In addition to estimating the required sample size, we may be interested in the power analysis for different sample and effect sizes. To do this in the G*Power, we follow the same steps as before and then we click on "X-Y plot for a range of values". We plot option "Power $(1 - \beta$ err prob)", as a function of "Total sample size" and choose 3 plots with effect sizes from 0.1 in steps of 0.15. **Figure 1b** shows the results of conducting the power analysis for the same example for various sample sizes and three effect sizes; low, medium and large effect sizes.

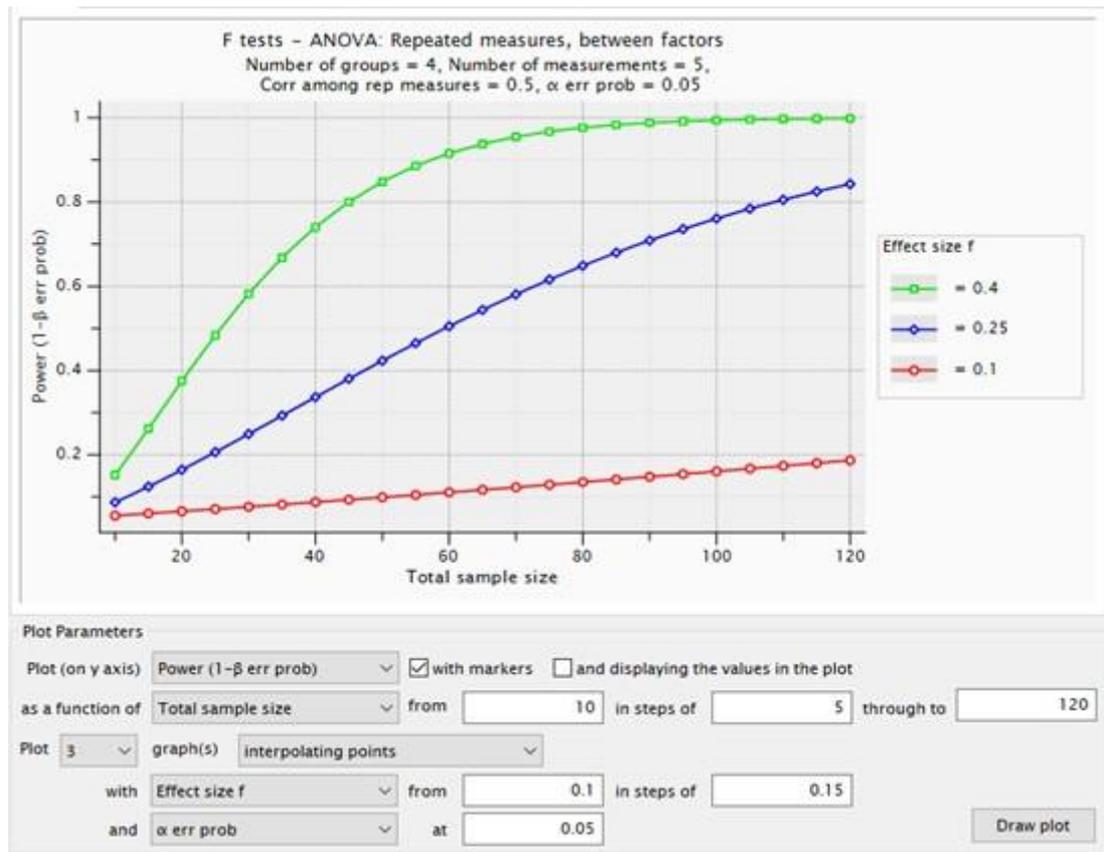

**Fig. 1b:** Power curves for various sample sizes and effect sizes using the G*Power for testing the differences among groups.

For the hypothesis testing in (S5), we also choose "F tests" in the test family button and then specify "ANOVA: Repeated measures, within factors" as the statistical test. To calculate the sample size, we select "A priori: Compute required sample size - given α, power, and effect size" as the type of power analysis. Then, for the input parameters, we assign $g = 4$ for the number of groups, $t = 5$ for the number of repeated measures, $\alpha = 0.05$ for the significance level, and $1 - \beta = 0.8$ for the desired statistical power. By following **Table 2**, we assign $f = 0.25$ as the medium effect size, $\rho = 0.5$ as the medium correlation among repeated measures, and $\varepsilon = 1$ as the nonsphericity correction. From the output results in **Figure 1c**, a sample size of 24 is necessary to provide a statistical power of 0.8 or above. The power curves associated with this test is depicted in **Figure 1d**.

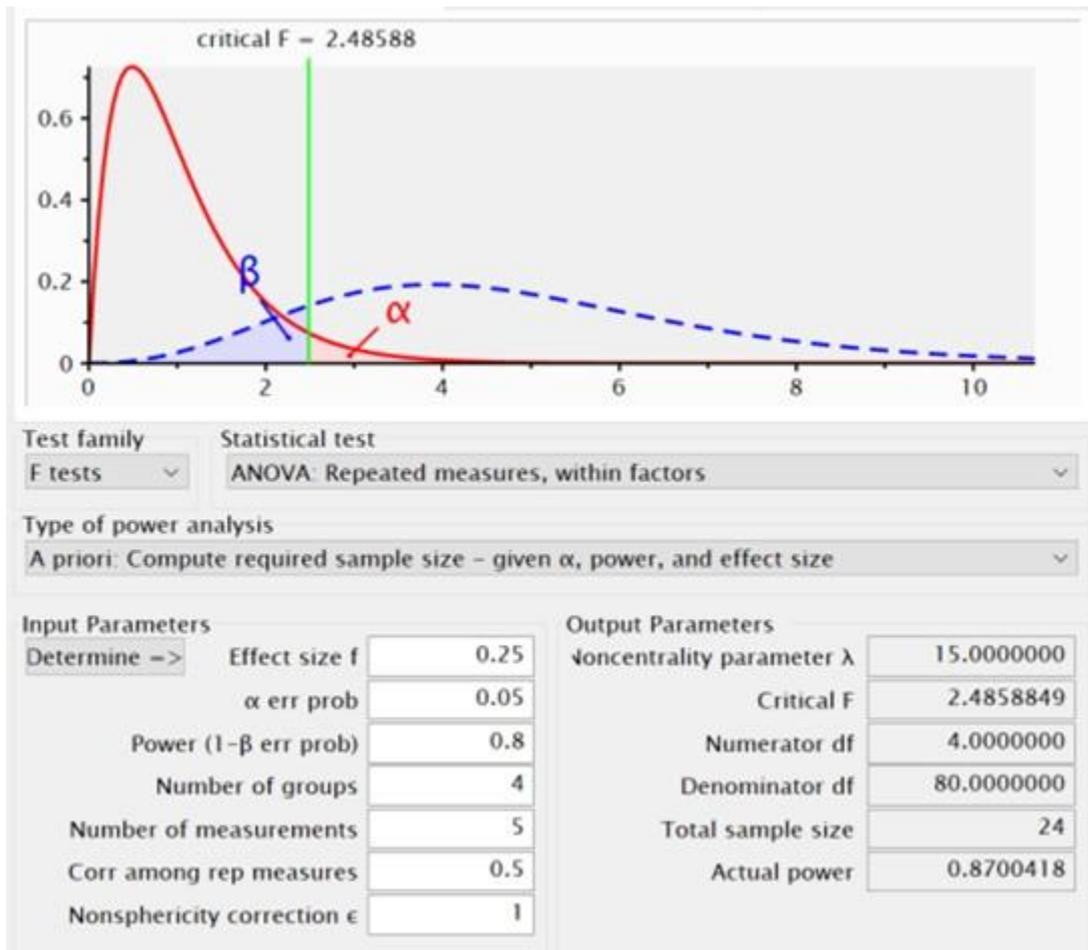

**Fig. 1c:** The a priori power analysis to calculate the sample size using the G*Power for testing the differences among time points.

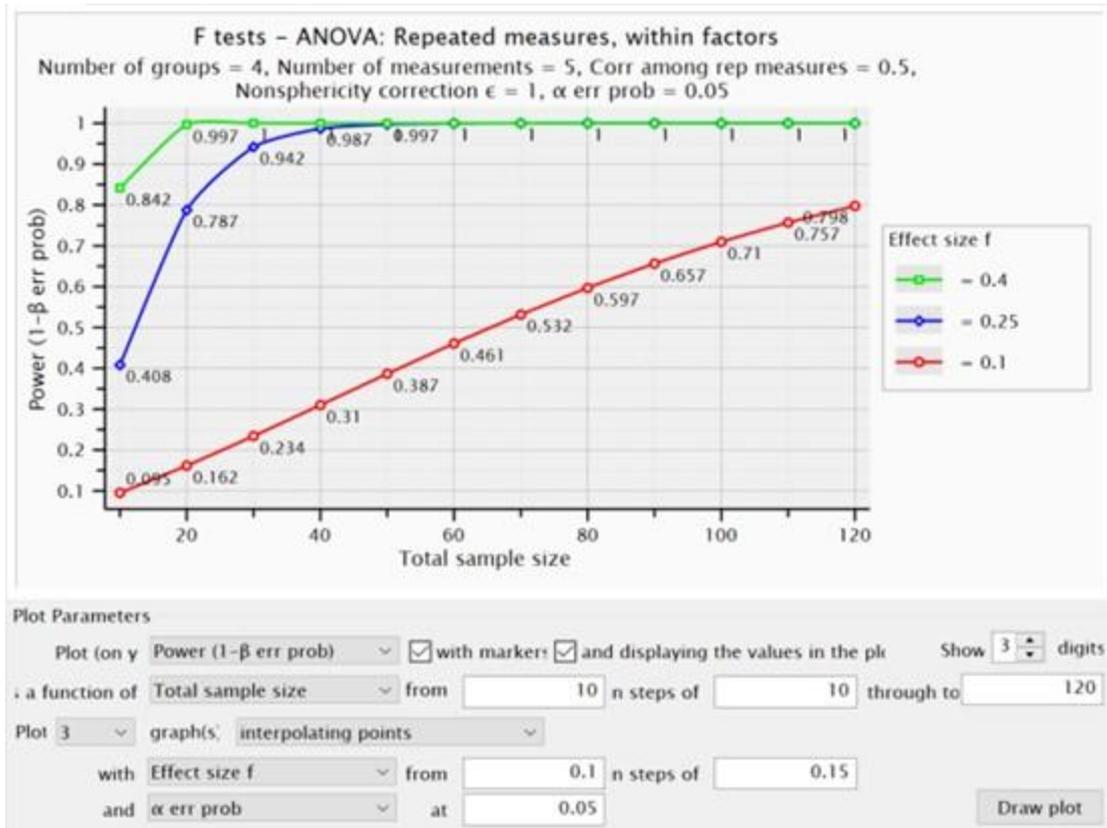

**Fig. 1d:** Power curves for various sample sizes and effect sizes using the G*Power for testing the differences among time points.

For the hypothesis testing in (S6), by setting "F tests" as the test family, we select "ANOVA: Repeated measures, within-between interaction" as the statistical test. To calculate the sample size, we choose "A priori: Compute required sample size - given α, power, and effect size" as the type of power analysis. For the input parameters, we assign $g = 4$ for the number of groups, $t = 5$ for the number of repeated measures, $\alpha = 0.05$ for the significance level, and $1 - \beta = 0.8$ as the desired statistical power. Based on the description in **Table 2**, we assign $f = 0.25$ as the medium effect size, $\rho = 0.5$ as the median correlation among repeated measures, and $\varepsilon = 1$ as the nonsphericity correction. The results from the G*Power are shown in **Figure 1e**, which indicates that it needs at least 32 samples to attain a statistical power of no less than 0.8. The power curves associated with this test is depicted in **Figure 1f**.

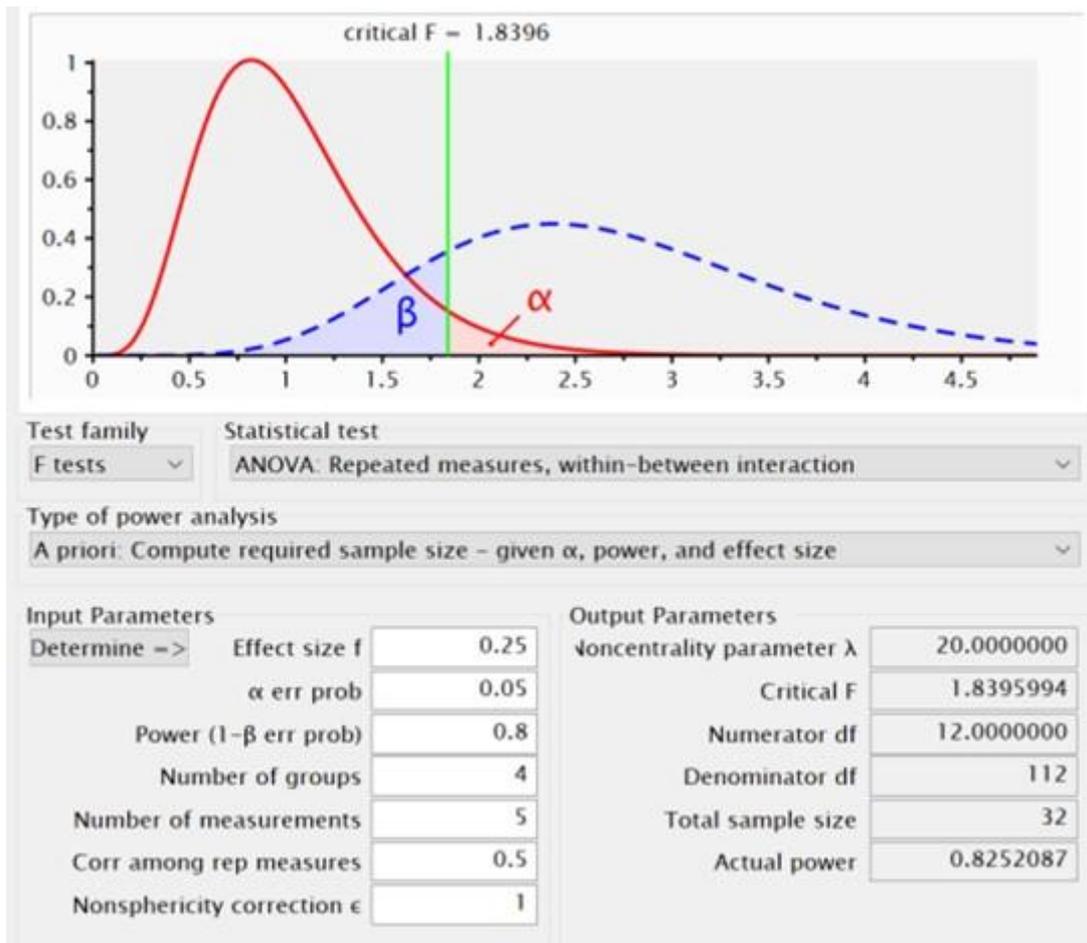

**Fig. 1e:** The a priori power analysis to calculate the sample size using the G*Power for testing the interaction between group and time.

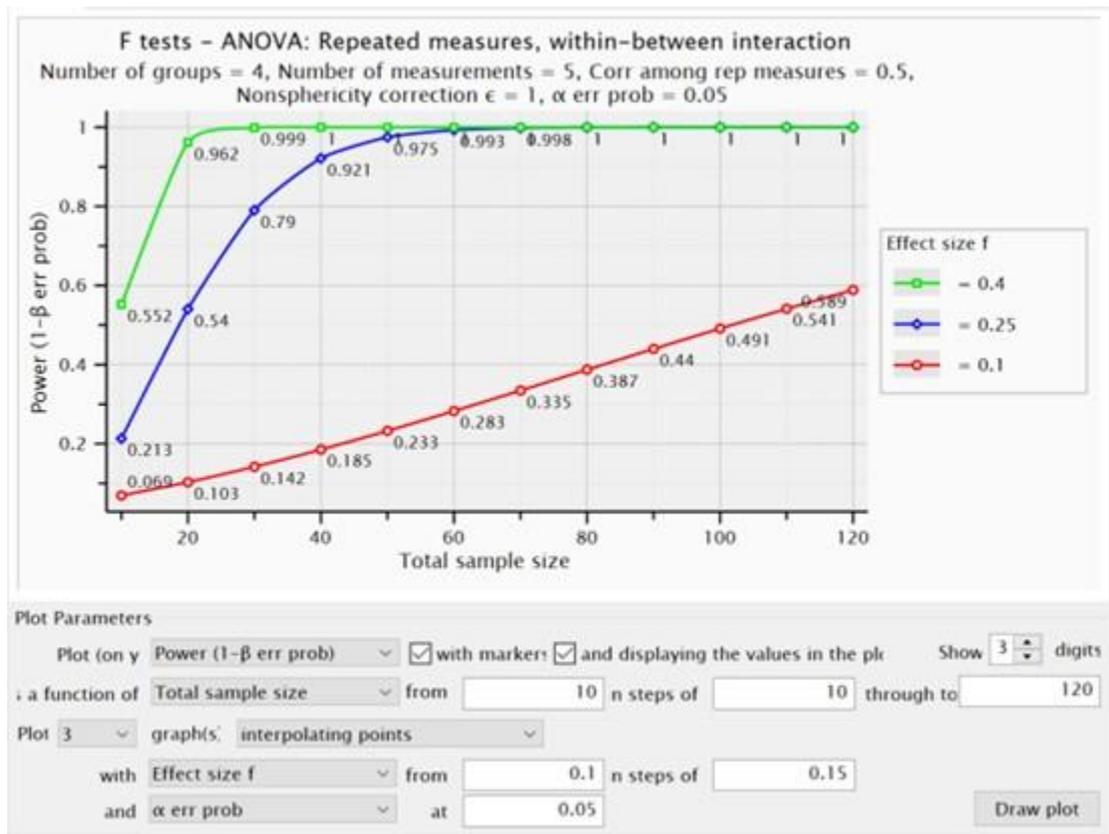

**Fig. 1f:** Power curves for various sample sizes and effect sizes using the G*Power for testing the interaction between group and time.

## V.     *In vivo* Experimental Design

The main objective of this study was to characterize hemi-transection (moderate to severe injury) and complete transection (very severe injury) spinal cord injury (SCI) in an *in vivo* rodent model. This injury model represents penetrating trauma (such as fracture or stab wound) and the progress of injury in rat closely mimics human pathophysiological events. We studied a total of 20 rats. Equal number of adult (~300 g) male and female rats were randomly divided into three injury and one control (laminectomy without injury) groups. Only the PI was aware of the group allocation at the different stages of the experiment. Each group (n=5 rats) underwent either hemi-transection (~1 mm clean transverse incision of exactly half of the spinal cord diameter) or the complete transection. General anesthesia was induced during all surgical procedures and neuro-electrophysiological recording sessions to avoid pain in animals. Laminectomy is a safe and secure procedure to expose the dorsal part of spinal cord.

Laminectomy can be performed on one or more vertebrae, and is a relatively easy procedure in rats, which does not cause stress, insult, or injury to the spinal cord parenchyma. Once the spinal cord in exposed, the injury will be performed under microscope, which may take a few short minutes. Then the paravertebral muscles are sutured back in order, skin closed, and disinfected. As shown below in **Figure 2**, the three experimental injury groups are: (i) left hemi-transection at T8, (ii) right hemi-transection at T10, and (iii) complete transection of the spinal cord at T8.

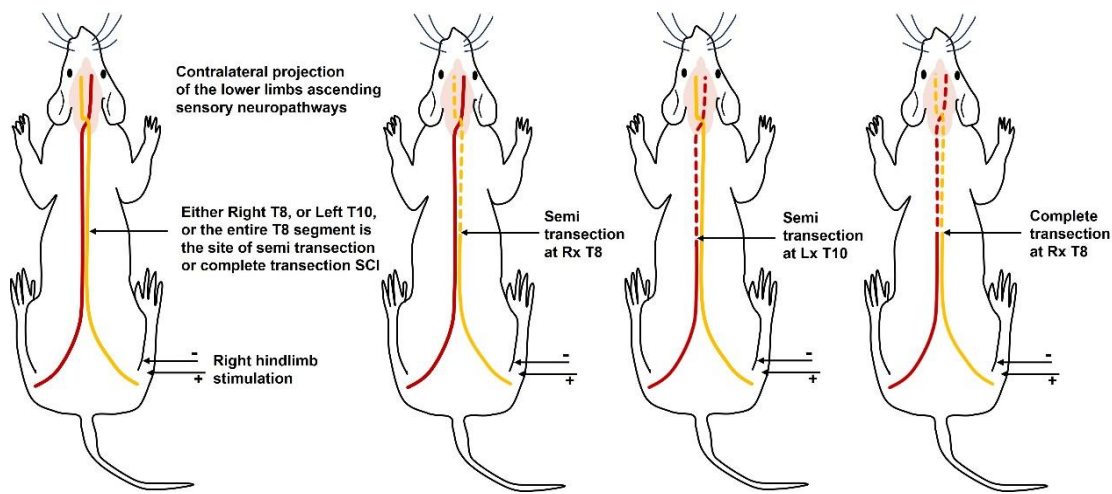

**Fig 2** shows a schematic of the three injury groups, electrical stimulation site, hindlimb ascending SSEP signals propagation, and decussation of sensory pathways to the contralateral cortices.

To be consistent among all subjects, only right hindlimb was stimulated (positive current pulses of 3.5 mA, 200 microseconds pulse width, and 1 Hz frequency) by inserting two needle electrodes near the Tibial nerve (but without touching the nerve bundle) of the right hindlimb. The somatosensory evoked potential (SSEP) signals are the central nervous system (brain) responses elicited by any form of sensory stimuli in the periphery. The SSEPs are recorded from the corresponding sensory area of the contralateral cortex (analog-to-digital conversion at a 5 kHz sampling rate and eliminating 50, 60, and 120 Hz frequencies interferences). After surgery, each rat received 5 ml intradermal saline twice a day for 2 days, 5 mg/kg *i.m.* Gentamicin for 4 days, and 0.3 mg/kg *s.c.* Buprenex analgesic for 3 days. Rats' bladder expressed manually twice a day, till spontaneous voiding observed. Rats had free access to food and water, which were placed on the bottom of their individual cages. All *in vivo* experimental procedures were approved by the Hong Kong Baptist University Institutional Animal Care and Use Committee ((23–139) in DH/HT&A/8/2/6 Pt.8) and followed the US-NIH guidelines for Neuroscience Research.

The SSEP signals were recorded (and normalized) after stimulating the right hindlimb on the day prior to injury (to obtain the baseline) and on days 4, 7, 14 and 21 post injury.

## VI.     Example Results

In this example, we have four experimental groups (one control and three injury groups:, left injury, right injury, and complete transection) with five SSEP recordings (baseline, days 4, 7, 14, and 21) to be analyzed. A sample size of five rats in each group was used. Based on this sample size with 80% statistical power, this can detect an effect size as low as 0.32 for a two-way repeated measure ANOVA, assuming a probability of type I error, $\alpha = 0.05$ and a medium level of between-subject correlation of 0.5.

**Single Group ANOVA**

**Table 3** tabulates the SSEP signals results after right hindlimb stimulation, T8 left hemi-transection injury, recorded from right cortex.

**Table 3:** Data collection design (one group, five subjects, five time points).

| Subjects (rats) | Baseline signals | Signals on day 4 | Signals on day 7 | Signals on day 14 | Signals on Day 21 |
|---|---|---|---|---|---|
| 1 | 1.0 | 1.2442 | 1.6132 | 1.2916 | 1.2916 |
| 2 | 1.0 | 0.964 | 1.8748 | 1.5555 | 1.5555 |
| 3 | 1.0 | 0.9318 | 0.8811 | 0.9753 | 0.5691 |
| 4 | 1.0 | 1.2502 | 1.1113 | 1.2712 | 0.7921 |
| 5 | 1.0 | 1.9395 | 1.4037 | 1.5248 | 0.9224 |

The null hypothesis is that the mean SSEP signals on days 4, 7, 14, and 21 are the same as the baseline. From data, we found the $F^*$ calculated to be 2.07 with p-value of 0.133. Therefore, we fail to reject the null hypothesis.

In conclusion, all rats with injury were paralyzed, and the severity of their injury depends on the severity and location of the transection injury. However, rats had no other complications. It was demonstrated that the neuro-electrophysiology monitoring, which is the only objective method to examine the onset, progress, and (any potential endogenous) recovery post-injury, clearly assessed the functional changes that are directly correlated to the severity of the injuries. As expected, stimulating the right hindlimb did not detect injury of the left hemi-transection injury. On the other hand, as described further and shown by the calculated *p-value* for **Table 3**, stimulating the right hindlimb will detect injury of the right hemi-transection injury.

**Multiple Groups Repeated Measures ANOVA**

Following the previously reported experimental example (**Table 3**), here we present three distinctive groups (**Table 4**). Group 1 represents subjects (five rats) with T8 left hemi-transection SCI. Group 2 presents subjects (other 5 rats) with T10 right hemi-transection injury. Group 3 presents data from 5 more rats with complete transection injury at T8 spinal cord. In all groups, the SSEP signals recorded from stimulating the right hindlimb and then, normalized. Similarly, there are one baseline SSEP monitoring (recorded 3 days before the injury) and four SSEP recordings on days 4, 7, 14, and 21 post-SCI.

**Table 4:** Data collection design (3 injury groups, 5 subjects/group, 5 time points).

| Experimental Injury Groups | Subjects (rats) | Baseline signal | Signal on day 4 | Signal on day 7 | Signal on day 14 | Signal on day 21 |
|---|---|---|---|---|---|---|
| Group 1 Left hemi-transection | 1 | 1 | 1.2442 | 1.6132 | 1.2916 | 1.2916 |
| | 2 | 1 | 0.964 | 1.8748 | 1.5555 | 1.5555 |
| | 3 | 1 | 0.9318 | 0.8811 | 0.9753 | 0.5691 |
| | 4 | 1 | 1.2502 | 1.1113 | 1.2712 | 0.7921 |
| | 5 | 1 | 1.9395 | 1.4037 | 1.5248 | 0.9224 |
| Group 2 Right hemi-transection | 1 | 1 | 1.1107 | 0.5734 | 0.7837 | 0.747 |
| | 2 | 1 | 0.9328 | 1.0583 | 0.4739 | 0.4917 |
| | 3 | 1 | 1.0061 | 1.4217 | 0.624 | 0.6284 |
| | 4 | 1 | 0.6884 | 0.9345 | 0.4775 | 0.7082 |
| | 5 | 1 | 0.1542 | 0.6776 | 0.4507 | 0.1745 |
| | 1 | 1 | 0.1989 | 0.193 | 1.1714 | 0.6429 |
| Group 3 Complete transection | 2 | 1 | 0.1593 | 0.1111 | 0.274 | 0.2158 |
| | 3 | 1 | 0.1013 | 0.3434 | 0.0867 | 0.0223 |
| | 4 | 1 | 0.1489 | 0.0604 | 0.0837 | 0.0678 |
| | 5 | 1 | 0.4105 | 0.2151 | 0.0001 | 0.0366 |

For the above example, we have $F^* = 25.785$ with *p-value* of 0.000 for the group effect, $F^* = 5.710$ with *p-value* of 0.001 for the time effect, and $F^* = 5.458$ with *p-value* of 0.000 for the interaction effect, respectively. These p-values were sufficient to reject the null hypotheses for the treatment effect, the group effect, and the interaction effect.

### VII. Discussion and conclusions

Sample size calculation is critical and has received substantial attention in biomedical research in the recent years. Today, power calculation and the sample size are considered to be two of the fundamental aspects of any *in vivo* research model [30-43]. Researchers noted that practically it would be impossible to obtain Institutional Animal Care and Use Committee (IACUC) approval, publish peer review papers, and be granted funding without incorporating description of well-designed experimental groups with detailed power and sample size calculation. In this paper, we have focused on studies with a number of repeated measures and provided a detailed explanation for sample size calculation in repeated measures ANOVA from the perspectives of both theory and application. We greatly recommend the G*Power for power analysis as it is user-friendly and efficient. We expect that this paper will aid researchers in conducting repeated measures ANOVA and calculate the sample size. Besides our work, we have also cited existing papers that provide a comprehensive review of power analysis for various statistical models.

Undoubtedly, studies with *in vivo* experimental procedures are a fundamental part of biomedical research and progress in healthcare. They are necessary to confidently bridge the gap between the results obtained from *in vitro* experiments, which are performed in a well-controlled microenvironment, and potentially clinical trials. Nevertheless, the ethical, moral, and legal rules (since 1966 when the "animal welfare act", which is the law that regulates the treatment of animals in research, was enacted) must be known, followed, and safeguarded. The rules to prevent undue suffering of animals in research could be summarized in three closely interlinked facts, so-called the three Biomedical 'R' rules. One, replace animals with alternative methods whenever it is possible. Two, reduce the number of animals to a minimal number that could provide significant and conclusive results (obviously lower than that number would be considered wasting resources since it will not be conclusive). And three, refine investigations to minimize their impacts on live animals (avoiding and preventing abuse of animals). In so doing, the following three aspects of any *in vivo* experimental design must be considered concurrently: (a) the power calculation, which is a statistical tool and a key process in conducting randomized control tests for computing the sample size and power, (b) the sample size, which is computed to determine the minimum number of animals per cohort that would offer meaningful outcome, and (c) the statistical analysis of each group, which is indispensable for decisively providing conclusive results. This paper aims to advocate for available tools that could be used enabling optimal execution of *in vivo* research investigations.

## Disclosure

**Author Contribution:** Angelo H. All (AH.A.) and Tiejun Tong (T.T.), contributed to the conception and experimental design. Hasan Al-Nashash (H.A.), Jiajin Wei (J.W.), Ke Yang (K.Y.), and Alzaatreh Ayman (A.A.) completed data analysis. AH.A., J.W., K.Y., A.A., and H.A. drafted the paper and performed literature search. AH.A., T.T., H.A. and Mohsen Adeli (M.A.) supervised and provided critical revisions. AH.A. and T.T. secured fundings.

**Data Archiving and Data Availability Statement:** The data reported in this manuscript from experimental and analytical studies are available from the corresponding author upon request.

**Ethical Statement:** We confirm that we followed all the institutional and governmental regulations concerning the ethical and legal issues governing biomedical research.

**Conflict of Interest:** All the authors declare that they have no conflict of interest.

**Acknowledgments:** Nil

**Source of Funding:** This study was supported by the Hong Kong Baptist University: Faculty Seed Fund # 31.4531.179234 (PI: A. H. All), Initiation Grant for Faculty Niche Research Areas (IG-FNRA) 2020/21 (PI: A. H. All), General Research Fund of Hong Kong (Project Number 12100121) 2021-22 (PI: A. H. All). The study was also supported by Dr. Tiejun Tong's General Research Fund of Hong Kong (Project Number HKBU12300123 and Project Number HKBU12303421) and the National Natural Science Foundation of China (Project Number 1207010822).

**References**

1. Kang H. Sample size determination and power analysis using the G*Power software. J. Educ. Eval. Health Prof. 2021;18:17. https://doi.org/10.3352/jeehp.2021.18.17.

2. Cohen J. Statistical Power Analysis for the Behavioral Sciences, 2nd ed. New York: Routledge; 1988.

3. Beck TW. The importance of a priori sample size estimation in strength and conditioning research. J. Strength Cond. Res. 2013;27(8):2323-37. 10.1519/JSC.0b013e318278eea0.

4. Perugini M, et al. A practical primer to power analysis for simple experimental designs. Int. Rev. Soc. Psychol. 2018;31:20. http://doi.org/10.5334/irsp.181.

5. Hertzog C, et al. Repeated measures analysis of variance in developmental research: selected issues. Child Dev. 1985;56:787-809. https://doi.org/10.2307/1130092.

6. Guo Y, et al. Selecting a sample size for studies with repeated measures. BMC Med. Res. Methodol. 2013;13:100. https://doi.org/10.1186/1471-2288-13-100.

7. All AH, et al. Comparative analysis of functional assessment for contusion and transection models of spinal cord injury. Spinal Cord. 2021;59(11):1206–9, https://doi.org/10.1038/s41393-021-00698-2.

8. Iyer S, et al. Multi-limb acquisition of motor evoked potentials and its application in spinal cord injury. J. Neurosci. Methods 2010;193(2):210–6. https://doi.org/10.1016/j.jneumeth.2010.08.017.

9. All AH, et al. Evoked potential and behavioral outcomes for experimental autoimmune encephalomyelitis in Lewis rats. Neurol. Sci. 2010;31(5):595–601. https://doi.org/10.1007/s10072-010-0329-y.

10. Agrawal G, et al Evoked potential versus behavior to detect minor insult to the spinal cord in a rat model. J. Clin. Neurosci. 2009;16(8):1052–5. https://doi.org/10.1016/j.jocn.2008.08.009.

11. All AH, et al. Effect of MOG sensitization on somatosensory evoked potential in Lewis rats. J. Neurol. Sci. 2009;284:81–9. https://doi.org/10.1016/j.jns.2009.04.025.

12. Bazley FA, et al. DTI for assessing axonal integrity after contusive spinal cord injury and transplantation of oligodendrocyte progenitor cells. Annu. Int. Conf. IEEE Eng. Med. Biol. Soc. 2012; pp 82–5. https://doi.org/10.1109/embc.2012.6345876.

13. Bazley FA, et al. The effects of local and general hypothermia on temperature profiles of the central nervous system following spinal cord injury in rats. Ther. Hypothermia Temp. Manag. 2014;4(3):115-24. https://doi.org/10.1089/ther.2014.0002.

14. All AH, et al. Characterization of transection spinal cord injuries by monitoring somatosensory evoked potentials and motor behavior. Brain Res. Bull. 2020;156:150–63.https://doi.org/10.1016/j.brainresbull.2019.12.012.

15. Bazley FA, et al. Enhancement of bilateral cortical somatosensory evoked potentials to intact forelimb stimulation following thoracic contusion spinal cord injury in rats. IEEE Trans. Neural Syst. Rehabilitation Eng. 2014;22(5):953–64. https://doi.org/10.1109/tnsre.2014.2319313.

16. Bazley FA, et al. Plasticity associated changes in cortical somatosensory evoked potentials following spinal cord injury in rats. Annu. Int. Conf. IEEE Eng. Med. Biol. Soc. 2011; pp 2005–8. https://doi.org/10.1109/iembs.2011.6090564.

17. Faul F, et al. G*Power 3: a flexible statistical power analysis program for the social, behavioral, and biomedical sciences. Behav. Res. Methods. 2007;39(2):175-91.

18. Scheffé H, The Analysis of Variance. New York: John Wiley & Sons; 1999.

19. Davis CS. Statistical Methods for the Analysis of Repeated Measurements. New York: Springer; 2002.

20. Kutner M H, et al. Applied linear statistical models, 5th ed. 2005, McGraw-Hill, New York.

21. Rana R K, et al. Analysis of repeated measurement data in the clinical trials, Journal of Ayurveda and Integrative Medicine, 2013;4:77-8. https://doi.org/10.4103/0975-9476.113872.

22. Keselman H J, et al. The analysis of repeated measures designs: a review, British Journal of Mathematical and Statistical Psychology, 2001;54:1-20. https://doi.org/10.1348/000711001159357

23. Quintana S M, et al. Monte Carlo Comparison of Seven ε-Adjustment Procedures in Repeated Measures Designs with Small Sample Sizes, Journal of Educational Statistics, 1994;19:57-71. https://doi.org/10.2307/1165177

24. Yazici B, et al. A comparison of various tests of normality, Journal of Statistical Computation and Simulation, 2007;77:175-83, https://doi.org/10.1080/10629360600678310

25. Sheldon M R, et al. The use and interpretation of the Friedman test in the analysis of ordinal-scale data in repeated measures designs, Physiotherapy Research International, 1996;1(4):221-8. https://doi.org/10.1002/pri.66.

26. SAS Library Repeated Measures ANOVA Using PROC GLM. https://stats.oarc.ucla.edu/sas/library/sas-libraryrepeated-measures-anova-using-sas-proc-glm/

27. Cohen J, Statistical power analysis for the behavioral sciences, 2013, Routledge.

28. Guo Y, et al. Selecting a sample size for studies with repeated measures. BMC Medical Research Methodology, 2013;13:1-8.

29. Kang H, Sample size determination and power analysis using the G* Power software. Journal of Educational Evaluation for Health Professions, 2021;18:17.

30. Agrawal G, et al. Slope analysis of somatosensory evoked potentials in spinal cord injury for detecting contusion injury and focal demyelination. Journal of Clinical Neuroscience 2010;17(9):1159-64. https://doi.org/10.1016/j.jocn.2010.02.005.

31. Bazley FA, et al. Direct Reprogramming of Human Primordial Germ Cells into Induced Pluripotent Stem Cells: Efficient Generation of Genetically Engineered Germ Cells. Stem Cells and Development 2015;24(22): 2634-48. https://doi.org/10.1089/scd.2015.0100.

32. Liang L, et al. Designing Upconversion Nanocrystals Capable of 745 nm Sensitization and 803 nm Emission for Deep-Tissue Imaging. Chemistry. 2016 ;22(31):10801-7. https://doi.org/10.1002/chem.201602514.

33. Al-Nashash H, et al. Spinal Cord Injury Detection and Monitoring Using Spectral Coherence. IEEE Transactions on Biomedical Engineering. 2009;56(8):1971-9. https://doi.org/10.1109/TBME.2009.2018296.

34. Pashai N, et al. Genome-wide profiling of pluripotent cells reveals a unique molecular signature of human embryonic germ cells. PLoS One. 2012;7(6):e39088. https://doi.org/10.1371/journal.pone.0039088.

35. Kadkhodaei M, et al. Automatic segmentation of multimodal brain tumor images based on classification of super-voxels. 2016 38th Annual International Conference of the IEEE Engineering in Medicine and Biology Society (EMBC), Orlando, FL, USA, 2016, pp. 5945-5948. https://doi.org/10.1109/EMBC.2016.7592082.

36. Agrawal G, et al. A comparative study of recording procedures for motor evoked potential signals. 2009 Annual International Conference of the IEEE Engineering in Medicine and Biology Society, Minneapolis, MN, USA, 2009, pp. 2086-2089. https://doi.org/10.1109/IEMBS.2009.5333953.

37. Mir H, et al. Novel Modeling of Somatosensory Evoked Potentials for the Assessment of Spinal Cord Injury. IEEE Transactions on Biomedical Engineering, 2018;65(3):511-20. https://doi.org/10.1109/TBME.2017.2700498.

38. Väyrynen E, et al. Automatic Parametrization of Somatosensory Evoked Potentials with Chirp Modeling. IEEE Transactions on Neural Systems and Rehabilitation Engineering, 2016;24(9):981-92. https://doi.org/10.1109/TNSRE.2016.2525829.

39. Mir H, et al. Spinal cord injury evaluation using morphological difference of somatosensory evoked potentials. 2011 5th International Conference on Bioinformatics and Biomedical Engineering, Wuhan, China, 2011, pp. 1-4. https://doi.org/10.1109/icbbe.2011.5780408.

40. Sherman DL, et al. Spinal cord integrity monitoring by adaptive coherence measurement. J Neurosci Methods. 2010;193(1):90-9. https://doi.org/10.1016/j.jneumeth.2010.07.035.

41. Mir HS, et al. Quantification of Spinal Cord Injury Level Using Somatosensory Evoked Potentials. 2010 4th International Conference on Bioinformatics and Biomedical Engineering, Chengdu, China, 2010, pp. 1-4. https://doi.org/10.1109/ICBBE.2010.5515579.

42. Agrawal G, et al. Shape analysis of Somatosensory Evoked Potentials to detect a focal spinal cord lesion. 2009 IEEE 35th Annual Northeast Bioengineering Conference, Cambridge, MA, USA, 2009, pp. 1-2. https://doi.org/10.1109/NEBC.2009.4967710.

43. Bazley FA, et al. A simple and effective semi-invasive method for inducing local hypothermia in rat spinal cord. 2013 35th Annual International Conference of the IEEE Engineering in Medicine and Biology Society (EMBC), Osaka, Japan, 2013, pp. 6321-6324. https://doi.org/10.1109/EMBC.2013.6610999.

# Supplementary

**Repeated Measures ANOVA Theory**

In a study with repeated measures, $g$ represents the number of groups, and $n_j$ represents the number of subjects in group $j$. This results in $n$, the total sample size, calculated as the sum of $n_j$ for all groups. Each subject undergoes repeated measurements at $t$ time points as depicted in Table 1. The response variable $y_{kij}$ is assumed to follow a normal distribution within each group.

The repeated measures ANOVA model for one sample problem (**Table 1** with $g = 1$) can be written as [20]:

$$y_{ij} = \mu_{..} + \rho_i + \tau_j + \varepsilon_{ij}, i = 1, \ldots, n, j = 1, \ldots, t, \tag{S1}$$

where $\mu_{..}$ is the grand mean, $\rho_i$ is the random effect of subject $i$, $\tau_j = \mu_{.j} - \mu_{..}$ is the fixed effect of time $j$ with $\sum_j \tau_j = 0$ and $\varepsilon_{ij}$ is the random error. The random variables $\varepsilon_{ij}$ and $\rho_i$ are assumed to be independent and follow normal distributions with fixed variances, since $g = 1$, $n_g = n$ and we drop the first indices of $y_{kij}$.

For testing the differences among the $t$ time points, the null and alternative hypotheses are:

$H_0$ : the mean responses are all equal among time points,

$H_1$ : the mean responses are not all equal among time points.

Based on the model in (S1), the above hypotheses can be translated as

$H_0: \tau_1 = \tau_2 = \cdots = \tau_t = 0$

$H_1: \tau_j \neq 0$ for some $j$ \hfill (S2)

The hypotheses in (S2) can be tested using $F$ test. $F^* = \frac{MS_{\text{treatment}}}{MS_{\text{error}}}$, where $MS_{\text{treatment}} = \sum_{i,j} \frac{(\bar{y}_{.j} - \bar{y}_{..})^2}{t-1}$ and $MS_{\text{error}} = \sum_{i,j} \frac{(y_{ij} - \bar{y}_{i.} - \bar{y}_{.j} - \bar{y}_{..})^2}{(n-1)(t-1)}$. Here, $\bar{y}_{.j} = \sum_i y_{ij}/n$ is the mean of subjects at time $j$ and $\bar{y}_{i.} = \sum_j y_{ij}/t$ is the mean of each subject over time. When the null hypothesis is true, the test statistic for (S2) follows an $F$ distribution with degrees of freedom in the numerator ($t$-1) and denominator ($n$-1)($t$-1). Accordingly, we reject the null hypothesis if the observed $F^*$ value is greater than the upper $\alpha$ quantile $F^{-1}_{t-1,(n-1)(t-1)}(1-\alpha)$, where $F^{-1}$ is the inverse function of the $F$ distribution cumulative function and $\alpha$ is the significance level. Note that $F^{-1}_{t-1,(n-1)(t-1)}(1-\alpha)$ is also known as the critical value of the test. Alternatively, the $p$-value of the test can be computed from $P(F > F^*)$ where $F$ follows an $F$ distribution with degrees of freedom in the numerator (*t*-1) and denominator (*n*-1)(*t*-1). $H_0$ is rejected if p-value$<\alpha$, and we will conclude that the mean responses are not all equal among time points.

**Assumptions of the Test**

Under the sphericity condition, the $F$ statistic follows $F_{t-1,(n-1)(t-1)}$ distribution, which can be tested using the Mauchly test [21]. It is noteworthy to mention that if the sphericity condition is not satisfied, using $F_{t-1,(n-1)(t-1)}$ may inflate the probability of type I error. In practice, the degree of freedoms of the $F$ test can be adjusted if the sphericity condition is not met. In this context, Greenhous and Geisser (GG) adjustment and Huynh and Feldt (HF) adjustment [22] are two typical examples, though other adjustments are also available in [23].

The repeated measures ANOVA is robust to the violation of the normality assumption for moderate sample size. However, in case of a small sample size, the violation of normality assumption may inflate the type I error. In practice, the normality can be checked using the Shapiro-Wilk test, the Kolmogorov-Smirnov test, and/or the normal probability plot [24]. If the normality assumption is violated, a transformation such as square root, logarithm, or other functions may be useful. In the event that the repeated measures ANOVA assumptions are violated, another option is to use the nonparametric analogue, the Friedman test [25], which requires less stringent assumptions.

**Multiple Samples Repeated Measures ANOVA Theory**

Multiple samples occur when there are more than one group ( $k > 1$ ). For example, when comparing the control group to the experimental group with repeated measurements are taken over time. For a multiple-sample problem, the repeated measures ANOVA model (**Table 1**) can be written as [20]

$$y_{kij} = \mu_{...} + \rho_{i(k)} + \tau_j + \gamma_k + (\tau\gamma)_{kj} + \varepsilon_{kij}, i = 1, ..., n_k, j = 1, ..., t, k = 1, ..., g, \quad (S3)$$

where $\mu_{...}$ is the grand mean, $\rho_{i(k)}$ is the random effect of subject $i$ in group $k$, $\tau_j = \mu_{..j} - \mu_{..}$ is the fixed effect of time $j$ with $\sum_j \tau_j = 0$, $\gamma_k = \mu_{k..} - \mu_{..}$ is the fixed effect of group $k$ with $\sum_k \gamma_k = 0$, $(\tau\gamma)_{kj} = \mu_{k.j} - \mu_{k..} - \mu_{..j} + \mu_{...}$ is the fixed effect of the interaction between time $j$ and group $k$ with $\sum_j (\tau\gamma)_{kj} = \sum_k (\tau\gamma)_{kj} = 0$. The random variables $\varepsilon_{kij}$ and $\rho_{i(k)}$ are assumed to be independent and follow normal distributions. Under the sphericity condition, $F$ tests can be used to test for group effects, time effects, or the interaction between group and time effects. Or on other dialect, three common questions to be addressed consist of: (i) whether the mean responses are significantly different among the $g$ groups; (ii) whether the mean responses are significantly different among the $t$ time points; and (iii) whether there exists a significant interaction between the two factors group and time. From the perspective of statistics, we can formulate them as three hypothesis testing problems as follows:

(i) For testing the differences among the $g$ groups, the null and alternative hypotheses are

$H_0$ : the mean responses are all equal among groups,

$H_1$ : the mean responses are not all equal among groups.

(ii) For testing the differences among the $t$ time points, the null and alternative hypotheses are

$H_0$ : the mean responses are all equal among time points,

$H_1$ : the mean responses are not all equal among time points.

(iii) For testing the interaction between group and time, the null and alternative hypotheses are

$H_0$ : there is no interaction between group and time,

$H_1$ : there is an interaction between group and time.

The $H_0$ and $H_1$ hypotheses reported in the multiple samples repeated measures ANOVA section can be translated based on the model in (S3) as follows.

For the group effects, we test the following hypotheses:

$$H_0: \gamma_1 = \gamma_2 = \cdots = \gamma_g = 0$$
$$H_1: \gamma_k \neq 0 \text{ for some } k. \tag{S4}$$

The test statistic of the hypotheses in (S4) is $F^* = \frac{MS_{group}}{MS_{subject(group)}}$, where $MS_{group} = \frac{\sum_{k,i,j}(\bar{y}_{k..}-\bar{y}_{...})^2}{g-1}$ and $MS_{subject(group)} = \frac{\sum_{k,i,j}(\bar{y}_{ki.}-\bar{y}_{k..})^2}{(n-g)}$. When the null hypothesis is true, the test statistic in (S4) follows an $F$ distribution with degrees of freedom $g-1$ and $n-g$. Accordingly, we reject the null hypothesis if the $F^*$ value is greater than the upper $\alpha$ quantile $F^{-1}_{g-1,n-g}(1-\alpha)$. The p-value can be computed from $P(F > F^*)$ where $F$ follows an $F$ distribution with degrees of freedom in the numerator (g-1) and denominator (n-g). $H_0$ is rejected if the p-value$<\alpha$.

For the time effects, we test the following hypotheses:

$$H_0: \tau_1 = \tau_2 = \cdots = \tau_t = 0$$
$$H_1: \tau_j \neq 0 \text{ for some } j. \tag{S5}$$

The test statistic of the hypotheses in (S5) is $F^* = \frac{MS_{treatment}}{MS_{error}}$, where $MS_{treatment} = \frac{\sum_{k,i,j}(\bar{y}_{..j}-\bar{y}_{...})^2}{t-1}$ and $MS_{error} = \sum_{k,i,j}\frac{(y_{kij}-\bar{y}_{k.j}-\bar{y}_{ki.}-\bar{y}_{k..})^2}{(n-g)(t-1)}$. Under the null hypothesis, the test statistic follows an $F$ distribution with degrees of freedom $t-1$ and $(n-g)(t-1)$. Accordingly, we reject the null hypothesis if the $F^*$ value is greater than the upper $\alpha$ quantile $F^{-1}_{t-1,(n-g)(t-1)}(1-\alpha)$. The p-value can be computed from $P(F > F^*)$ where $F$ follows an $F$ distribution with degrees of freedom in the numerator (t-1) and denominator (n-g)(t-1). Reject $H_0$ if p-value$<\alpha$.

For the group × time interaction effects, we test the following hypotheses:

$$H_0: (\tau\gamma)_{11} = (\tau\gamma)_{12} = (\tau\gamma)_{13} = \cdots = (\tau\gamma)_{gt} = 0$$
$$H_1: (\tau\gamma)_{kj} \neq 0 \text{ for some } k \text{ and } j \tag{S6}$$

The test statistic of the hypotheses in (S6) is $F^* = \frac{MS_{group \times treatment}}{MS_{error}}$, where

$$MS_{\text{group}\times\text{treatment}} = \frac{\sum_{k,i,j}(\bar{y}_{k.j}-\bar{y}_{k..}-\bar{y}_{..j}+\bar{y}_{...})^2}{(g-1)(t-1)} \text{ and } MS_{error} = \frac{\sum_{k,i,j}(y_{kij}-\bar{y}_{k.j}-\bar{y}_{ki.}+\bar{y}_{k..})^2}{(n-g)(t-1)}.$$

The dot notation in $\bar{y}_{kij}$ represents the mean over the dot index. For example, $\bar{y}_{k..}$ is the mean of group $k$. Under the null hypothesis, the test statistic for (S6) follows an $F$ distribution with degrees of freedom $(g-1)(t-1)$ and $(n-g)(t-1)$. Accordingly, we reject the null hypothesis if the $F^*$ value is greater than the upper $\alpha$ quantile $F^{-1}_{(g-1)(t-1),(n-g)(t-1)}(1-\alpha)$. The *p*-value and can be computed from $P(F > F^*)$ where $F$ follows an $F$ distribution with degrees of freedom in the numerator (g-1)(*t*-1) and denominator (*n-g*)(*t*-1). Reject $H_0$ if p-value$<\alpha$.